\documentclass[12pt,runningheads]{fotfs}

\usepackage{amsmath, amssymb, graphics, latexsym, makeidx}
\usepackage[all,dvips]{xy}
\usepackage{times}

\newcommand
{\thla}{\ifmmode\hbox{\boldmath\hbox{$\twoheadleftarrow$}}\else$\hbox{\boldm
ath\hbox{$\twoheadleftarrow$}}$\fi}
\newcommand
{\thlavar}[1]{\ifmmode\hbox{\boldmath\hbox{$\twoheadleftarrow$}}\hbox{\scriptsize 
\hbox{$#1$}}\else$\hbox{\boldmath\hbox{$\twoheadleftarrow$}}\hbox{\scriptsize 
\hbox{$#1$}}$\fi}
\newcommand
{\thlagenvar}[1]{\ifmmode\hbox{\boldmath\hbox{$\twoheadleftarrow$}}\hbox{\sc
riptsize\hbox{($#1$)}}\else$\hbox{\boldmath\hbox{$\twoheadleftarrow$}}\hbox{
\scriptsize\hbox{($#1$)}}$\fi}
\newcommand {\tildavar}[2]{\ifmmode{#1}^{\thla #2}\else${#1}^{\thla #2}$\fi}

\makeindex
\begin{document}
\frontmatter

\mainmatter


\title{An  $\omega$-power of a  context-free language \\  which is Borel above
$\Delta^0_\omega$}
\titlerunning{A  Borel  $\omega$-power of a  context-free language   above
$\Delta^0_\omega$}

\author{Jacques Duparc and Olivier Finkel}

\institute{Universit\'e de Lausanne,\\ 
Information Systems Institute, and\\
Western Swiss Center for Logic, History and Philosophy of Sciences \\ 
B\^atiment Provence\\
CH-1015  Lausanne\\
\\
and 
\\
\\{\it Equipe Mod\`{e}les de Calcul et Complexit\'e}  
 \\ {\it Laboratoire de l'Informatique du Parall\'elisme}
\footnote{UMR 5668 - CNRS - ENS Lyon - UCB Lyon - INRIA \\ LIP Research Report RR 2007-17} \\  CNRS et Ecole Normale Sup\'erieure de Lyon
 \\ 46, All\'ee d'Italie 69364 Lyon Cedex 07, France.  }

\email{jacques.duparc@unil.ch and Olivier.Finkel@ens-lyon.fr}

\received{...}
\revised{...}
\accepted{...}

\subjclass{{\bf PRIMARY} SECONDARY}

\maketitle
\begin{abstract}
We use erasers-like basic operations on words to construct
a set that is both Borel and above $\Delta^0_\omega$, built as a set $V^\omega$  
where $V$ is a language of finite words accepted by a pushdown automaton.  In particular, this gives a first  example of an
$\omega$-power of a context free language which is a Borel set 
 of  infinite rank.

\end{abstract}


\section{Preliminaries}
\label{preliminaries}

Given a set $A$ (called the alphabet) we write $A^*$,
and $A^\omega$, for the sets of finite, and infinite words over
$A$. We denote the empty word by $\epsilon$.
In order to facilitate the reading, we use $u,v,w$ for finite words,
and $x,y,z$ for infinite words. Given two words $u$ and
$v$ (respectively, $u$ and $y$), we write ${uv}$ (respectively,${uy}$)
 for the concatenation of $u$ and $v$ (respectively, of $u$ and $y$).
Let $U \subseteq A^*$ and $Y \subseteq A^*\cup A^{\omega}$, we set:
${UY}=\{ uv,  uy: u\in X \wedge v, y \in Y \}$.

We recall that, given a language $V\subseteq A^*$,
the $\omega$-{\em power} of this language is
$$V^\omega=\{x=u_1u_2\ldots u_n\ldots\in A^\omega:\ \forall n<\omega\ u_n
\in V\setminus\{\varepsilon\}\}$$

$A^\omega$ is equipped with the usual topology. i.e. the product of the discrete
topology on the alphabet $A$. So that every open set is of the form
$WA^\omega$
for any $W\subseteq A^*$. Or, to say it differently, every closed set is
defined as the set of
all infinite branches of a tree over $A$. We work within the Borel
hierarchy
of sets which is the strictly increasing (for inclusion) sequence of
classes of sets $(\Sigma^0_\xi)_{\xi<\omega_1}$ - together with the dual
classes $(\Pi^0_\xi)_{\xi<\omega_1}$ and the ambiguous ones
$(\Delta^0_\xi)_{\xi<\omega_1}$ - which reports how many operations of
countable unions and intersections are necessary to produce a Borel set on
the basis of the open ones.

A reduction relation between sets $X,\ Y$ is a partial ordering $X\leq Y$ which expresses that the problem of knowing whether any element $x$ belongs to
$X$
is at most as complicated as deciding whether $f(x)$ belongs to $Y$, for
some given 
{\em simple} function $f$. A very natural reduction relation between sets of
infinite words (closely related to reals), has been thoroughly studied by Wadge in the seventies.
From the topological point of view,
{\em simple} means continuous, therefore the Wadge ordering compares sets of infinite sequences with respect to their fine topological complexity. Associated with determinacy, this partial
ordering becomes a pre-wellordering with anti-chains of length at most two.
The so called Wadge Hierarchy it induces incredibly refines the old Borel
Hierarchy. Determinacy makes it way through a representation of 
continuous functions in terms of strategies for player II in a suitable two-player
game: the Wadge game $W(X,Y)$. In this game, players I and II, take turn playing letters
of 
the alphabet corresponding to $X$ for I, and letters of the alphabet
corresponding to $Y$ for II. In order to get the right correspondence between a strategy for player II and a continuous
function, player II is allowed to skip, whereas I is not. However, II must play infinitely many letters.

As usual, reduction relations induce the notion of a complete set: a set that both belongs to some class, whose members it also reduces. In the context of Wadge reducibility, a set is complete if it belongs to some class closed
by inverse image of continuous functions, and reduces everyone
of its members. A class which admits a complete set is called a Wadge Class.
As
a matter of fact, all $\Sigma^0_\xi$, and $\Pi^0_\xi$, are Wadge classes, whereas $\Delta^0_\xi$ ($\xi>1$) are not.

For instance, the set of all infinite sequences that contains a $1$ is
$\Sigma^0_1$-complete, the one that contains infinitely many
$1$s is $\Pi^0_2$-complete. As a matter of fact, reaching complete sets for upper levels of the Borel
hierarchy, requires other means which we introduce in next sections.

\section{Erasers}

For climbing up along the finite levels of the Borel hierarchy, we use erasers-like moves, see 
\cite{Dup1}. For simplicity, imagine a player (either I or II) playing 
a Wadge game, in charge of a set $X\subseteq {A}^\omega$, with the
extra possibility to
delete any terminal part of her last moves.

We recall the definition of the operation $X \mapsto X^\approx$ over
sets of
infinite words. It was first introduced in \cite{Fin01a} by the second
author, and is a simple variant of the first author's operation of
exponentiation
$X \mapsto X^\sim$ which first appeared in \cite{Dup1}.

 We denote $|v|$ the length of any finite word $v$. 
 If  $|v|=0$, $v$ is the empty word. If $v=v_1v_2\ldots v_k$ where $k\geq 1$ and each $v_i$ is in $A$, then 
$|v|=k$ and we write $v(i)=v_i$  and $v[i]=v(1)\ldots v(i)$ for $i\leq k$ ; so $v[0]=\epsilon$.
 The prefix relation is denoted $\sqsubseteq$: the finite word $u$ is a prefix of the finite 
word $v$
(denoted $u\sqsubseteq v$) if and only if there exists a (finite) word $w$ such that $v=uw$.
 the finite word $u$ is a prefix of the $\omega$-word $x$ (denoted $u\sqsubseteq x$) 
iff there exists an $\omega$-word $y$ such that $x=uy$.

Given a finite alphabet $A$, 
we write $A^{\leq \omega}$ for $A^* \cup A^\omega$. 

\begin{definition}
Let $A$ be any finite alphabet, $\thla ~ \notin A$,  $B=A \cup \{ \thla \}$, and $x\in B^{\leq\omega}$, then 
\newline $x^{\thla}$ is inductively defined by:
\newline $\epsilon^{\thla} =\epsilon$, and for a finite word $u\in (A\cup \{\thla\})^*$: 
\newline $(ua)^{\thla}=u^{\thla}a$, if $a\in A$,
\newline $(u\thla)^{\thla} =u^{\thla}$  with its  last letter removed if $|u^{\thla}|>0$, 
\newline $(u\thla)^{\thla}$ is undefined if $|u^{\thla}|=0$,
\newline and for $u$ infinite:
\newline $(u)^{\thla} = \lim_{n\in\omega} (u[n])^{\thla}$, where, given $\beta_n$ and $v$ in   $A^*$,
\newline $v\sqsubseteq \lim_{n\in\omega} \beta_n \leftrightarrow  \exists n \forall p\geq n\quad  \beta_p[|v|]=v$.
\end{definition}

We now make easy this definition to understand by describing it informally. 
For $x \in B^{\leq \omega}$, $x^{\thla}$
denotes the string $x$, once every $\thla$
occurring in $x$
has been ``evaluated" to the back space operation (the one familiar to your
computer!),
proceeding from left to right inside $x$. In other words $x^{\thla} = x$ from
which every
interval of the form $`` a\thla "$ ($a\in A$) is removed. By convention, we
assume
$(u\thla)^{\thla}$ is undefined when $u^{\thla}$ is the empty sequence. i.e. when the last
letter
$\thla$ cannot be used
as an eraser (because every letter of $A$ in $u$
has already been erased by some eraser $\thla$ placed in $u$). 
We remark that the resulting word $x^{\thla}$ may be
finite or infinite. 

For instance,\begin{itemize}
\item if $u=(a\thla)^n$, for $n\geq 1$, or
$u=(a\thla)^\omega$ then $(u)^{\thla}=\epsilon$,

\item if $u=(ab\thla)^\omega$ then $(u)^{\thla}=a^\omega$,

\item if $u=bb(\thla a)^\omega$ then $(u)^{\thla}=b$,
\item if $u=\thla(a\thla)^\omega$ or $u=a\thla\thla a^\omega$ or
$u=(a\thla\thla)^\omega$
then $(u)^{\thla}$ is undefined.
\end{itemize}

\begin{definition}
For $X\subseteq A^{\omega}$, ~~~~
$$X^\approx =\{x\in (A\cup \{\thla\})^{\omega}:\ x^{\thla}\in X\}.$$
\end{definition}

\noindent The following result easily follows from \cite{Dup1} and was
applied
in \cite{Fin01a,Fin04} to study the $\omega$-powers of finitary context free
languages.

\begin{theorem}\label{thedup}
Let $n$ be an integer $\geq 2$ and $X\subseteq A^\omega$ be a ${\bf
\Pi_n^0}$-complete
set. Then $X^\approx$ is a
${\bf \Pi_{n+1}^0}$-complete subset of $(A\cup\{\thla\})^\omega$.
\end{theorem}

Next remarks will be essential later.

\begin{remark}\label{remarkfunction1} Consider the following function:
$$f:\ x\in (A\cup \{\thla\})^{\omega}\mapsto y\in A^\omega$$
defined by:
\begin{itemize}
\item $y= 0^\omega$ if $x^{\thla}$ is finite or undefined,
\item $y= x^{\thla}$ otherwise.

\end{itemize}

It is clearly Borel. In fact a quick computation shows that the inverse
image
of any basic clopen set is Borel of low finite rank.
\end{remark}

\begin{remark} Let $X$ be any subset of the Cantor space $\{0, 1\}^\omega$, and $f$ as in
remark
\ref{remarkfunction1}.
If $0^\omega\not\in X$, then for any $x\in\{0,1,\thla\}^\omega$
$$x\in X^{\approx} \iff f(x)\in X$$
In other words, $X^{\approx}=f^{-1}X$. In particular, if $X$ is Borel, so is
$X^{\approx}$
\end{remark}

\section{Increasing sequences of erasers}

\noindent  The following construction has been partly  used by the second author in \cite{Fin04} to construct a Borel 
set of infinite rank which is an $\omega$-power, i.e. in the form $V^\omega$, where $V$ is a set 
of finite words over a finite alphabet $\Sigma$.     
   We iterate the operation   $X \mapsto X^\approx$ finitely many  times, and take the limit.
More precisely,

\begin{definition}
Given any set $X\subseteq A^{\omega}$:   \begin{itemize}
\item $X_k^{\approx 0}=X$,
\item $X_k^{\approx 1}=X^\approx$,
\item $X_k^{\approx 2}=(X_k^{\approx 1})^\approx$,
\item $X_k^{\approx (k)}=(X_k^{\approx (k-1)})^\approx$, where we apply $k$
times the operation $X\mapsto X^\approx$
with different new letters  $\thla_k$,  $\thla_{k-1}$,   \ldots ,    $\thla_{2}$,    $\thla_{1}$,   

in such a way that we have
successively:\begin{itemize}

\item $X_k^{\approx 0}=X\subseteq A^{\omega}$,
\item $X_k^{\approx 1}\subseteq (A\cup\{\thla_k\})^{\omega}$,
\item $X_k^{\approx 2} \subseteq (A\cup\{\thla_k,  \thla_{k-1}\})^{\omega}$,
\item $X_k^{\approx (k)} \subseteq
(A \cup\{\thla_1, \thla_{2}, \ldots , \thla_k\})^{\omega}$.
\end{itemize}
\item We set $X^{\approx (k)} = X_k^{\approx (k)}$.  

\end{itemize}
$X^{\approx \infty} \subseteq (A\cup \{\thla_{n}:\ 0<n<\omega\})^{\omega}$
is
defined by $$x\in X^{\approx \infty}\iff_{def}$$\begin{itemize}

\item $\mbox{for each integer }n,
\ x_n={{{{x^{\thla_1}}^{\ldots}}^{\thla_{n-1}}}^{\thla_n}}$ is defined,
infinite, and
\item $x_\infty=\lim_{n<\omega}x_n$ is defined,
infinite, and belongs to $X$. \end{itemize}

\end{definition}

\begin{remark}\label{remarkfunction} Consider the following sequence of
functions: \begin{itemize}
\item $f_0(x)=x$ ($f_0$ is the identity),

\item $f_{k+1}:\ (A\cup \{\thla_{n}:\ k<n<\omega\})^{\omega}\longmapsto
(A\cup \{\thla_{n}:\ k+1<n<\omega\})^{\omega}$

defined by:
\begin{itemize}
\item $f_{k+1}(x)= x^{\thla_{k+1}}$ if $x^{\thla_{k+1}}$ is infinite,
\item $f_{k+1}(x)= 0^\omega$ if $x^{\thla_{k+1}}$ is finite or undefined,
\end{itemize}

\end{itemize}

By induction on $k$, one shows that every function $f_k$ is Borel - and even
Borel of finite rank.

Moreover, since Borel functions are closed under taking the limits \cite{Kur}, the following function is Borel.
$$f_{\infty}:\ (A\cup \{\thla_{n}:\
0<n<\omega\})^{\omega}\longmapsto
A^{\omega}$$
defined by:
\begin{itemize}
\item $f_{\infty}(x)= \lim_{n<\omega} f_n({x})$ if $\lim_{n<\omega} f_n({x})$
is
defined, and infinite,
\item $f_{\infty}(x)= 0^\omega$ otherwise.
\end{itemize}

\end{remark}

\begin{remark}\label{borel} Let $X\subset \{0,1\}^\omega$ with
$0^\omega\not\in X$, then for any
$$x\in(\{0,1\}\cup \{\thla_{n}:\ 0<n<\omega\})^{\omega}$$
$$x\in X^{\approx\infty} \iff f_\infty(x)\in X$$
In other words, $ X^{\approx\infty}={f_\infty}^{-1}(X)$, which shows that
whenever $X$ is Borel, $X^{\approx\infty}$ is Borel too.

\end{remark}

In fact, with tools described in \cite{Dup1}, and \cite{Dup2}, it is possible to show that given any $\Pi^0_{1}$-complete set  $Y$, the set 
$Y^{\approx\infty}$ belongs to $\Pi^0_{\omega+2}$. 
\newline If $X$ is the set of infinite words over the alphabet $\{0, 1\}$ which contains 
an infinite number of $1$s, then it is also possible to show that
 $X^{\approx\infty}$ is
Borel by completely different methods involving
decompositions of $\omega$-powers \cite{finsim,Fin04}.

\begin{proposition}
\label{notdelta}
Let $X$ be the set of infinite words over $\{0,1\}$ that contain infinitely
many $1$s,

$$X^{\approx\infty}\in \Delta^1_1\setminus\Delta^0_{\omega}$$

\end{proposition}

\begin{proof} The fact $X^{\approx\infty}$ is Borel is Remark \ref{borel}.
As for $X^{\approx\infty}\notin\Delta^0_{\omega}$, it
is a consequence of the fact that the operation $Y\longmapsto Y^{\approx}$
is
strictly increasing (for the Wadge ordering) inside $\Delta^0_{\omega}$
(see \cite{Dup1}\cite{Dup2}).
In other words, for any $Y\in\Delta^0_{\omega}$ the relation
$Y<_W Y^{\approx}$ holds ($<_W$ stands for the
strict Wadge ordering). But, as a matter of fact,
$\left(X^{\approx\infty}\right)^{\approx}\leq_W X^{\approx\infty}$
holds which forbids $X^{\approx\infty}$ to belong to
$\Delta^0_{\omega}$.

Indeed, to see that $\left(X^{\approx\infty}\right)^{\approx}\leq_W
X^{\approx\infty}$ holds,
it is enough to describe a winning strategy for player II in the
Wadge game $W\left(\left(X^{\approx\infty}\right)^{\approx},
X^{\approx\infty}\right)$. In this game, player II uses $\omega$ many different 
erasers:
$\thla_1, \thla_2, \thla_3,\ldots$ whose strength is opposite to their
indices ($\thla_k$ erases all erasers $\thla_j$ for any $j>k$ but no $\thla_i$ for $i\leq k$). While player I uses the same erasers as player II does,  plus
an
extra one ($\thla$) which is stronger than all the other ones.

The winning 
strategy for II derives from ordinal arithmetic: $1+\omega=\omega$. It
consists in copying I's run with a shift on the indices of erasers:
\begin{itemize}
\item if I plays a letter $0$ or $1$, then II plays the same letter,
\item if I plays an eraser $\thlavar{n}$, II plays the eraser
$\thlavar{n+1}$.
\item if I plays the eraser $\thla$ (the first one that will be taken into
account when the erasing process starts), then II plays $\thlavar{0}$.
\end{itemize}
This strategy is clearly winning.
\end{proof}

\section{Simulating $X^{\approx\infty}$ by the $\omega$-power of a
context-free
language}

It was already known that there exists an $\omega$-power of a finitary language which is 
Borel of infinite rank \cite{Fin04}. But the question was left open whether such a finitary language could be {\it  context free}.

This article provides effectively a context free language $V$  such that 
$V^\omega$ is a Borel set of infinite rank, and uses infinite Wadge games to show that this $\omega$-power 
$V^\omega$ is located above ${\bf \Delta}^0_\omega$ in the Borel hierarchy. 

The idea is to have $X^{\approx\infty}$, where $X$ stands for the set of all 
infinite words over $\{0,1\}$ that contain infinitely many $1$s to
be of the form $V^\omega$ for some language $V$ recognized by a
(non deterministic) Pushdown Automaton.   
We first recall the notion of pushdown automaton \cite{Berstel79,ABB96}.

\begin{definition} A pushdown automaton (PDA) is a 7-tuple
$$M=(Q,A,\Gamma, \delta, q_0, Z_0, F)$$
 where\begin{itemize}
\item $Q$ is a finite set of states,
\item $A$ is a finite input alphabet,
\item $\Gamma$ is a finite pushdown alphabet,
\item $q_0\in Q$ is the initial state, $Z_0 \in\Gamma$ is the start symbol,
\item $\delta$ is a mapping from $Q \times (A\cup\{\varepsilon\})\times
\Gamma$
to finite subsets of $Q\times \Gamma^*$.
\item $F\subseteq Q$ is the set of final states.
\end{itemize}

If $\gamma\in\Gamma^{+}$ describes the pushdown store content,
the leftmost symbol of $\gamma$ will be assumed to be on ``top" of the store.
A configuration of a PDA  is a pair $(q, \gamma)$ where $q\in Q$ and
$\gamma\in\Gamma^*$.

For $a\in A\cup\{\varepsilon\}$, $\gamma,\beta\in\Gamma^{*}$
and $Z\in\Gamma$, if $(p,\beta)$ is in $\delta(q,a,Z)$, then we write
$a: (q,Z\gamma)\mapsto_M (p,\beta\gamma)$.

$\mapsto_M^*$ is the transitive and reflexive closure of $\mapsto_M$.

Let $u =a_1a_2\ldots a_n$ be a finite word over $A$.
A finite sequence of configurations $r=(q_i,\gamma_i)_{1\leq i\leq p}$ is called
a run of $M$ on $u$, starting in configuration $(p,\gamma)$, iff:
\begin{enumerate}
\item $(q_1,\gamma_1)=(p,\gamma)$

\item for each $i$, $1\leq i \leq p-1$, there exists $b_i\in A\cup\{\varepsilon\}$
satisfying $b_i: (q_i,\gamma_i)\mapsto_M(q_{i+1},\gamma_{i+1})$
such that       ~ $a_1a_2\ldots a_n =b_1b_2\ldots b_{p-1}$. 
\end{enumerate}
This  run is simply called a run of  $M$ on $u$ if it  starts 
 from configuration $(q_0, Z_0)$.    

The language accepted by $M$ is\begin{center}
$L(M)= \{u\in A^*$:\ there is a  run $r$
of $M$ on $u$ ending in a final state$\}$.
\end{center}
\end{definition}

For instance, the set $0^*1\subset \{0,1\}^*$
is trivially context-free.
\begin{proposition}[Finkel] Let $L_n\mbox{ be the maximal subset of }$

$\{0,1,\thla_1,\thla_2,
\ldots,\thla_n\}^*\mbox{ such that }{{{{L_n^{\thla_1}}^{\thla_{2}}}^{\ldots}}^{\thla_n}}= 0^*1$,
$$L_n\mbox{ is context-free}$$
\end{proposition}
This was first noticed by the second author in \cite{Fin01a}.

To be more precise, by $u\in L_n$ we mean: 
we start with some $u$, then we evaluate $\thla_1$ as an eraser, and obtain $u_1$ (providing that we must never use $\thla_1$ to erase the empty sequence, i.e.
every
occurrence of a $\thla_1$ symbol does erase a letter $0$ or $1$ or an eraser
$\thla_i$ for $i>1$). Then we start again with $u_1$, this time we evaluate $\thla_2$ as an eraser, which yields $u_2$, and so on. When there is no more symbol $\thla_i$ to be evaluated, we are left with $u_n\in \{0,1\}^*$. We define $u\in L_n$ iff $u_n\in 0^*1$.

To make a PDA recognize
$L_n$, the idea is to have it guess (non
deterministically), 
for each single letter that it reads, whether this letter will be erased later or not.
Moreover, the PDA should also guess for each eraser it encounters, whether this eraser should
be
used as an eraser or whether it should not - for the only reason that it will be erased
later on by a {\em stronger} eraser. During the reading, the stack should be used to accumulate all pendant guesses, in order to verify later on that they are fulfilled.

We would very much like to prove that $L_\infty={\displaystyle
\bigcup_{n<\omega}L_n}$ is context-free. Unfortunately, we cannot get such a result. However, we
are able to show that a slightly more complicated set (strictly containing $L_\infty$)
is indeed context-free.

Of course, the first problem that comes to mind when working with $L_\infty$, is to
handle $\omega$ many different erasers with a finite alphabet. This implies that erasers
must be coded by finite words.
This was done by the second author in \cite{Fin03b}. Roughly speaking, the
eraser
$\thla_n$ is coded by the word $\alpha B^n C^n D^n E^n \beta$ with new letters $\alpha, B, C, D, E, \beta$. It is a little bit tricky, but
the PDA must really be able to read 
the number $n$ identifying the eraser four times.

The very definition of the sets $L_n$, requires the
erasing operations to be executed in an increasing order: in a word
that contains only the erasers $\thla_1,\ldots,
\thla_n$, one must consider first the eraser $\thla_1$, then $\thla_{2}$,
and
so on\ldots

Therefore this erasing process satisfy the following properties:
\begin{enumerate}
\item[(a)] An eraser $\thla_j$ may only erase letters $c\in \{0,1\}$ or
erasers $\thla_k$ with $k>j$.

\item[(b)] Assume that in a word $u \in L_n$, there is
a sequence
$cvw$ where $c$ is either in $\{0,1\}$ or in the set $\{\thla_1, \ldots,
\thla_{n-1}\}$,
and $w$ is (the code of) an eraser $\thla_k$ which erases $c$ once the erasing process 
is achieved. If there is in $v$ (the code of)
an eraser $\thla_j$ which erases $e$, where $e\in \{0,1\}$ or $e$
is (the code of)
another eraser, then $e$ must belong to $v$
(it is between $c$ and $w$ in the word $u$) ; moreover the erasing - by the eraser $\thla_j$ - has been achieved before the other one
with the eraser
$\thla_k$. This implies $j \leq k$. Thus the integer $k$ must
satisfy:
$$k \geq \max \{j:\mbox{ an eraser } \thla_j \mbox{ was used
inside } v\}$$

\end{enumerate}

The essential difference with the case studied in \cite{Fin03b} is that here 
an eraser $ \thla_j$ may only erase letters $0$ or $1$ or erasers $ \thla_k$ 
for $k>j$,  while in \cite{Fin03b} an eraser  $ \thla_j$ was assumed to be only 
able to  erase letters $0$ or $1$ or erasers $ \thla_k$ 
for $k<j$. So the above inequality was replaced by:
$$k \leq \min \{j:\mbox{ an eraser } \thla_j \mbox{ was used
inside } v\}$$

  However,  with a slight modification,   
 we can construct  a PDA $\mathcal{B}$ which, among 
words where letters  $\alpha, \beta,    B, C, D, E$ are only  
used to code erasers of the form $ \thla_j$, accepts 
exactly the words which belong to the language $L_\infty$.  
We now explain  the behavior of this PDA. 
(For simplicity,  we sometimes talk about the eraser $\thla_j$ instead of its code $\alpha B^jC^jD^jE^j\beta$.)

\vspace{5mm} Assume that $\mathcal{A}$ is a finite automaton accepting (by final state) 
the finitary language $0^*1$ over the alphabet $A=\{0, 1\}$. 
\newline  We can informally describe the behavior of the PDA  $\mathcal{B}$ when reading 
a word $u$ such that the letters $\alpha, B, C, D, E, \beta$  
are only used in $u$ to code the erasers 
$\thla_j$ for $1\leq j$. 

$\mathcal{B}$ 
simulates the  automaton $\mathcal{A}$ until it guesses (non 
deterministically) that it begins to read a segment 
$w$ which contains erasers which really erase and some letters of $A$ or some other 
erasers which are erased when the operations 
of erasing are achieved in $u$. 

 Then, still non deterministically, when $\mathcal{B}$ reads 
a letter $c\in A$
it may  guess that this letter will be erased and push it in the pushdown store, 
keeping in memory the current state of the automaton $\mathcal{A}$. 

 In a similar manner when   $\mathcal{B}$ reads 
the  code $\thla_j=\alpha B^jC^jD^jE^j\beta$, it may guess that 
this eraser will be erased (by another eraser $\thla_k$ with $k<j$) and then may push  
in the store the finite word $\gamma E^j \nu$, where $\gamma$, $E$, $\nu$ 
are in the pushdown alphabet of $\mathcal{B}$.

  But  $\mathcal{B}$ 
may also guess that the eraser $\thla_j=\alpha B^jC^jD^jE^j\beta$ will really be used
as an eraser.  If it guesses that the code of $\thla_j$ will be used as an eraser, 
$\mathcal{B}$ has to pop from the top of the pushdown store either a letter  $c\in A$ or 
the code $\gamma E^i.\nu$ of another eraser $\thla_i$, with $i>j$, which 
is erased by $\thla_j$. 

In this case, it is easy for $\mathcal{B}$ to check whether $i>j$ when reading the initial segment 
$\alpha B^j$ of $\thla_j$. 

 But as we remarked in $(b)$, the PDA $\mathcal{B}$ must also check 
that the integer $j$ is greater than or equal to every integer $p$ such that 
an eraser $\thla_p$ has been used since the letter $c\in A$ or the code 
 $\gamma E^i.\nu$ was pushed 
in the store.  Then, after having pushed some letter $t \in A$ 
or the code $t=\gamma E^i.\nu$ of an eraser in the pushdown store, and before popping it from the top of the 
stack, $\mathcal{B}$ must keep track of the following integer in the memory stack.

$$k = max [ p ~/~ \mbox{ some eraser } \thla_p  \mbox{ has been used since  } t
 \mbox{ was pushed in the stack }]$$

\noindent For that purpose $\mathcal{B}$ pushes the finite word $L_2 S^k L_1$ in the 
pushdown store ($L_1$ is pushed first, then $S^k$ and the letter $L_2$), with $L_1, L_2$ and $S$ are new letters added to the pushdown 
alphabet. 

 So, when $\mathcal{B}$ guesses that $\thla_j=\alpha B^jC^jD^jE^j\beta$ will be 
really used as an eraser, there is on top of the stack either a letter $c\in A$ 
or a code $\gamma E^i.\nu$ 
of an eraser which will be erased or a code $L_2 S^k L_1$. The 
 behavior of $\mathcal{B}$ is then as follows. 

  Assume first  there is a code $L_2 S^k L_1$ on top of the stack.  
Then $\mathcal{B}$ firstly checks that $j\geq k$ holds by reading the 
segment $\alpha B^j C$ of the eraser $\alpha B^jC^jD^jE^j\beta$.

If $j \geq k$ holds, then using $\epsilon$-transitions, $\mathcal{B}$ completely pops the word  $L_2 S^k L_1$ from the top of the stack.
($\mathcal{B}$ has already checked it is allowed to use the eraser $\thla_j$). 

 Then, in each case, the top of the stack contains either a letter $c\in A$, 
or the code $\gamma E^i\nu$ 
of an eraser which should be erased later. 
$\mathcal{B}$ pops this letter $c$ or the code 
$\gamma E^i.\nu$ (having checked that $j < i$ after reading the segment 
$\alpha B^j C^j$ of the eraser $\alpha B^jC^jD^jE^j\beta$).

A this point, we must have a look at the top stack symbols. There are three  cases: 
\begin{enumerate}

\item The top stack symbol is the bottom symbol $Z_0$. In which case, the PDA $\mathcal{B}$, after having completely read 
the eraser $\thla_j$, may pursue the simulation of the automaton  $\mathcal{A}$ 
or guess that it begins to read another segment $v$ which will be erased. Hence  
the next letter $c\in A$ 
or the next code $\alpha B^m.C^m.D^m.E^m.\beta$ of the word will be erased. Then 
$\mathcal{B}$ pushes the letter $c\in A$ or the code $\gamma E^m.\nu$ 
of $\thla_m$ in the pushdown store.

\item If the top stack symbol is either 
 a letter $c'\in A$ or a code $\gamma E^m.\nu$, then  $\mathcal{B}$ pushes 
the code $L_2 S^j L_1$ in the pushdown store ( $j$ is then the maximum of the set of 
integers $p$ such that an eraser $\thla_p$ has been used since the letter $c'$ or 
the code $\gamma E^m.\nu$ has been pushed into the stack). 
 
\item If the top stack symbols are a code $L_2 S^l L_1$, then the PDA 
$\mathcal{B}$ must compare the integers $j$ and $l$, and replace 
$L_2 S^l L_1$ by $L_2 S^j L_1$ in case $j>l$. $\mathcal{B}$ achieves this task  while reading 
the segment $D^j E^j \beta$ of the eraser $\alpha B^jC^jD^jE^j\beta$.
\newline The PDA $\mathcal{B}$ pops a letter $S$ for each letter $D$ it reads. Then it checks whether $j \geq l$ is satisfied. 
\newline If $j \geq l$ then it pushes $L_2 S^j L_1$ while it reads the 
segment $E^j \beta$ of the eraser $\thla_j$.
\newline In case $j < l$, after it reads $D^j$, the part $S^{l-j} L_1$ of the code $L_2 S^l L_1$ remains in the stack. The PDA then pushes again $j$ letters 
$S$ and a letter $L_2$ while  reading $E^j \beta$. 
\end{enumerate}

\noindent  When again the stack only contains $Z_0$ - the initial stack symbol - $\mathcal{B}$ resumes the simulation of the automaton $\mathcal{A}$ or it 
guesses that it begins to read a new segment which will be erased later.

\vspace{5mm} We are confronted with the fact $\mathcal{B}$  will also accept some words where the letters 
$\alpha, \beta, B, C, D, E$ are not used to code erasers. How can we make sure that this PDA is not misled by such wrong codes of erasers ?

\section{Wrong codes of erasers and the right $\omega$-power}

In fact, one cannot make sure that a PDA notices the discrepancy between
right codes of the form $\alpha B^jC^jD^jE^j\beta$ and wrong ones (of the
form $\alpha B^bC^cD^dE^e\beta$ where $b,c,d,e$ are
not all the same integer for instance).
However, there is a satisfactory solution: instead of
having a PDA reject these wrong codes, simply let it accept all of them.
Accepting a word if it contains a wrong code of an eraser is trivial for a non deterministic 
PDA. So instead of a PDA $\mathcal{B}$ that accepts precisely
$L_\infty$
(up to the coding of erasers), we set
\begin{proposition} There exists a PDA $\mathcal{B}$ s.t.
$$\mathcal{L}(\mathcal{B})=L_\infty \cup W$$

\end{proposition}
where $W$ stands for the set of all finite words which host a wrong code,
$L_\infty$ really is $L_\infty$ where erasers are replaced by their correct
codes, and $\mathcal{L}(\mathcal{B})$ is the language recognized by
$\mathcal{B}$. 
Everything is ready for the main result.

\begin{theorem}
\label{setrecognized}
The $\omega$-power $Y=\mathcal{L}(\mathcal{B})^\omega$  of the context-free language ${L}(\mathcal{B})$
 described above satisfies

$$Y\in\Delta^1_1\smallsetminus\Delta^0_{\omega}$$

\end{theorem}

\begin{proof} To begin with, the set $Y$ is the disjoint union of three
different sets: $Y=Y_0\cup Y_\infty\cup Y_{*}$, where $Y_0$ is the set of all infinite
sequences in $Y$  with no wrong code in them, $Y_\infty$ the set of all
infinite sequences with infinitely many wrong codes, and $Y_*$ the
set of infinite sequences with finitely many wrong codes
(at least one). We remark that:
\begin{itemize}
\item $Y_0$ is Wadge equivalent to the set $X^{\approx{\infty}}$ as defined
in
\ref{notdelta}. i.e. the set of all $\omega$-words that,
after taking care of the erasing process, ultimately reduce to words with
infinitely many $1$s. To be more precise,
it is this very same set up to a renaming of the erasers. So $Y_0$ belongs
to
$\Delta^1_1$.

\item $Y_\infty$ is Wadge equivalent to $X$, so it is $\Pi_2^0$-complete.

\item $Y_*$ is more complicated. However, it is of the form $Y_*=WY_0$,
where
$W$ is the set of all finite words with at least an occurrence of
a
wrong code. So $Y_*$ is a countable union of sets, each of which is Wadge
equivalent to $Y_0$. Hence, $Y_*$ is a countable union of
Borel sets, therefore $Y_*$ is Borel too.
\end{itemize}
All three cases put together show that $Y=Y_0\cup Y_\infty\cup Y_{*}$, is a
finite union of Borel sets, hence it Borel too.

It remains to prove that $Y\notin\Delta^0_\omega$. This, in fact, is
immediate
from Proposition \ref{notdelta} which stated that
$X^{\approx{\infty}}\notin\Delta^0_\omega$. Because 
there is an obvious winning strategy for player II in
the Wadge game $W(X^{\approx{\infty}},Y)$. It consists in never playing a wrong
code, and copying I's run up to the renaming of the erasers. Since
$X^{\approx{\infty}}$ is clearly Wadge equivalent to $Y_0$ this strategy
works perfectly well and shows that $Y\notin\Delta^0_\omega$.

\end{proof}

This quick study gives an example of how an infinite
game theoretical approach leads to intriguing results in Theoretical
Computer
Science. On one hand,
the notion of erasers is highly related to the dynamic behavior of
players in games. And, on the other hand, non determinism
provides very effective ways to deal with the erasing process. So, all
together, they afford a method for describing (topological) complexity of
very effective sets of reals.

\vspace{5mm}
\noindent
{\bf  Acknowledgments.}   we wish to thank an anonymous referee for useful comments on 
a previous version of this paper.
\articleend

\backmatter
\newpage
\pagestyle{empty}
\printindex
\newpage
\pagestyle{empty}



\end{document}